\documentclass[iop]{emulateapj}
\usepackage{rotating}
\usepackage{mathtools}
\usepackage{graphicx}
\usepackage{multirow}

\slugcomment{}

\shorttitle{}
\shortauthors{Pace \& Salim}

\begin{document}

\title{Satellites of radio AGN in SDSS: insights into AGN triggering and feedback}

\author{Cameron Pace \& Samir Salim}
\affil{Indiana University, Dept. of Astronomy, Swain Hall West 319, Bloomington, IN, USA 47405-7105}
\email{cjpace@indiana.edu, salims@indiana.edu}

\begin{abstract}
We study the effects of radio jets on galaxies in their vicinity (satellites) and the role of satellites in
triggering radio-loud active galactic nuclei (AGNs). The study compares the aggregate properties of satellites
of a sample of 7,220 radio AGNs at $z < 0.3$ (identified by Best \& Heckman 2012 from the SDSS
and NVSS+FIRST surveys) to the satellites of a control sample of radio-quiet galaxies, which are
matched in redshift, color, luminosity, and axis ratio, as well as by environment type: field galaxies,
cluster members and brightest cluster galaxies (BCGs). Remarkably, we find that radio AGNs exhibit on average a 50\% excess (17$\sigma$ significance) in the number of satellites within 100 kpc even though the cluster membership was controlled for (e.g., radio BCGs have more satellites than radio-quiet BCGs, etc.).  Satellite excess is not confirmed for high-excitation sources, which are only 2\% of radio AGN. Extra satellites may be responsible for raising the probability for hot gas AGN accretion via tidal effects or may otherwise enhance the intensity or duration of the radio-emitting phase. Furthermore, we find that the incidence of radio AGNs among potential hosts (massive ellipticals) is similar for field galaxies and for non-BCG cluster members, suggesting that AGN fueling depends primarily on conditions in the host halo rather than the parent, cluster halo. Regarding feedback, we find that radio AGNs, either high or low excitation, have no detectable effect on star formation in their satellites, as neither induced star formation nor star formation quenching is present in more than $\sim 1\%$ of radio AGN.
\end{abstract}

\keywords{galaxies: active, galaxies: interactions, galaxies: jets}

\section{Introduction}

\label{sec:intro}

%Active galactic nuclei (AGN) are hosted by galaxies that span a range of morphological types, but the hosts of radio-loud AGN (R-AGN) are generally massive ellipticals (Best at al. 2005, Kauffmann et al. 2008). 
Active galactic nuclei (AGN)  are believed to be powered by two fundamentally different modes of accretion. Optically selected and luminous  radio-loud AGN (R-AGN) are powered by the radiatively-efficient accretion of gas onto a central supermassive black hole (SMBH) at a rate of one to ten percent of the Eddington limit. In contrast the majority of R-AGN, which have a low radio luminosity, accrete gas in a radiatively inefficient manner at a rate below one percent of the Eddington limit \citep{2012MNRAS.421.1569B}. The mode of accretion may also be related to galaxy morphology. While AGN in general are hosted by galaxies that span a range of morphological types,  the hosts of R-AGN are most often massive ellipticals
(e.g., Ekers \& Ekers 1973, Best at al. 2005b, Kauffmann et al. 2008).

This difference in R-AGN accretion processes could be a result of differences in the fueling mechanism. 
Optical and luminous radio AGN may be triggered by mergers and one-on-one interactions (Heckman et al. 1986, Barnes \& Hernquist 1996) which could supply the large quantities of cold gas required to maintain a high accretion rate. Low luminosity R-AGN may be triggered by the accretion of hot gas from the surrounding halo (Fabian 1994, Allen et al. 2006), although there could be additional triggering mechanisms \citep{2013MNRAS.430..638S}.  The key to understanding these triggering mechanisms therefore lies in the kpc to Mpc-scale environments of R-AGN, which is populated by satellite galaxies. 

Many studies have examined the relationship between R-AGN incidence and the environment.
  \cite{2007MNRAS.379..894B} found that R-AGN are more frequently found in central group and cluster galaxies when compared to galaxies of similar stellar mass.
 \cite{2004MNRAS.351...70B} studied a small sample of 91 R-AGN and found that the fraction of R-AGN shows little dependence on the Mpc-scale local galaxy density, but is dependent on the number of galaxies in the group in which the R-AGN is located, while the slightly larger sample of 212 R-AGN of \cite{2006ApJ...650..717R} showed a slight increase in R-AGN incidence with density at the Mpc scale. \cite{2005MNRAS.362...25B} and \cite{2013MNRAS.430..638S} both found a strong dependence of R-AGN activity on the stellar mass of the host galaxy, but when this is accounted for, R-AGN activity still has a dependence on density at the Mpc scale. \cite{2008MNRAS.384..953K} found that the environments of R-AGN at the $\sim$100 kpc scale are roughly twice as dense as those of radio-quiet AGN or radio-quiet galaxies of similar mass. In contrast to these studies, \cite{2013arXiv1305.2673W} found that at the $\sim$100 kpc scale only powerful R-AGN have higher clustering than radio-quiet galaxies of the same mass, and that for scales greater than about 160 kpc the clustering of R-AGN is similar to that of radio-quiet galaxies. 

The aforementioned studies find an increase in the R-AGN incidence rate in denser, cluster environments. This should not be surprising considering that R-AGN likely feed on the hot gas that is plentiful in clusters. However, it must be acknowledged that neither all ellipticals or brightest cluster galaxies (BCGs) are radio loud, nor are all massive field ellipticals radio quiet. The approach taken in the current study is therefore different as it separates the general environment (BCG, cluster member, or field galaxy) from the small-scale environment ($<100$ kpc). We compare the small-scale environments  (i.e. the satellite populations) of a statistically
large sample of R-AGN and a matching control sample of radio-quiet galaxies. Control galaxies are chosen not only to match the mass, type, and star formation history of an R-AGN but also the large scale
environment it is found in, such that the controls of field R-AGN are also field galaxies, the controls of cluster member R-AGN are also cluster members, and the controls of BCG R-AGN are also BCGs. 
This careful matching allows us to compare the satellite populations of R-AGN and otherwise similar radio-quiet galaxies to determine why only a subset of galaxies with similar environments become R-AGN. 

In our study of satellite populations we will separately focus on radio AGN with high accretion rates.
 The two modes of R-AGN accretion can be distinguished with emission line ratios \citep{1994ASPC...54..201L}.  
 \cite{2010A&A...509A...6B} defined an `excitation index' composed of four optical emission line ratios which can be used to separate R-AGN with a low accretion rate, which are known as low excitation radio galaxies (LERGs) from those with a high accretion rate, known as high excitation radio galaxies (HERGs). At most radio luminosities HERGs comprise only a few percent of the total R-AGN population, although they become more common at the highest radio luminosities \citep{2012MNRAS.421.1569B}. Many studies employ the Fanaroff-Riley classification, which classifies R-AGN based on their radio morphology \citep{1974MNRAS.167P..31F}. 
  Luminous R-AGN, which tend to be edge brightened, are known as Fanaroff-Riley type 2 (FR2) sources, while core-brightened lower-luminosity sources are referred to as Fanaroff-Riley type 1 (FR1) sources. Although LERGs and HERGs often exhibit FR1 and FR2 morphology, respectively,  \cite{1994ASPC...54..201L} demonstrated that a number of FR2 sources have low-excitation spectra (i.e. are LERGs). Since the LERG \& HERG classifications are more closely related to accretion mode and do not require high resolution radio maps, we adopt their use in our study, as determined by \cite{2012MNRAS.421.1569B}.
 
In addition to exploring the role of satellites in R-AGN triggering, the second major goal of this study is to understand if and to what extent satellites can be affected through feedback from the powerful jets that are the hallmark of R-AGN. These highly collimated jets originate in the nucleus of the AGN at relativistic speeds and may extend for kiloparsecs or even megaparsecs beyond the host galaxy (Tremblay et al. 2010, Schoenmakers et al. 2000). Radio AGN deposit most of their energy into the interstellar or intergalactic medium kinetically via their high-velocity jets (see Fabian 2012 for a recent review). Such interactions may be responsible for quenching star formation in the host galaxy, and there is some evidence that R-AGN may lead to quenching of star formation in their satellite galaxies.  \cite{2011MNRAS.413.2815S} found that satellite galaxies in the projected jet paths of FR2 sources are redder than satellites outside the jet path, but no such trend was found for the satellites of FR1 sources. This suggests that FR2 jets tend to quench star formation in satellite galaxies while FR1 jets do not. We will revisit these results in our study by using a much larger sample of R-AGN and by contrasting entire satellite populations of R-AGN (regardless of proximity to the jet) to those of the radio-quiet control sample. 

Somewhat paradoxically, radio jets are also considered as the mechanism behind possible \textit{positive} feedback, both in the host galaxy and in adjacent satellite galaxies. The idea is that interactions with jets may induce star formation by driving shocks into dense clouds which then collapse \citep{2008MNRAS.389.1750A}. Such AGN-driven star formation may have been important in building up the massive spheroid in the early phases of galaxy formation \citep{2009ApJ...700..262S}. The radio jets of both HERGs and LERGs are believed to be capable of producing positive feedback outside of the host as well. The powerful radio jets associated with HERGs pierce external clouds in the IGM, but triggered star formation may proceed in their slowly expanding radio lobes (De Young 1981, Bicknell et al. 2000). Since the jets of LERGs are generally less powerful, they may be capable of inducing star formation even in head-on collisions \citep{2004IAUS..222..485V}. 

A few candidates of such jet-induced star formation in the vicinity of R-AGN have been observed. The star forming region `09.6' in the eastern lobe of the nearby FR 2 source 3C 285 was first observed by \cite{1993ApJ...414..563V}. \emph{Chandra} observations confirm that this region is indeed experiencing a starburst phase \citep{2007ApJ...662..166H}, while a satellite galaxy of the FR2 source PKS2250--41, which is not currently in the projected jet path, also shows evidence of jet-induced star formation \citep{2008MNRAS.386.1797I}. 

 Induced star formation as a result of interactions with weak R-AGN has also been observed: one example is the LERG Centaurus A.
 \cite{2012MNRAS.421.1603C} recently showed that the youngest stars in the inner filament of Centaurus A are only a few Myrs old and are probably the result of the shock induced collapse of a molecular cloud.  Downstream from the inner star-forming filament, the radio jet interacts with an H I cloud and produces another filament of recently-formed stars \citep{1998ApJ...502..245G}.
 Minkowski's Object (MO) is another example of LERG-induced star formation. This star forming region is 15 kpc away from its host R-AGN NGC 541.  Strong UV and H$\alpha$ emission are indicative of the starburst nature of this object \citep{2004IAUS..222..485V}. Simulations  by \cite{2004ApJ...604...74F} have reproduced the observational characteristics of MO, including the star formation rate (SFR) of 0.3 M$_\sun$yr$^{-1}$. While it is possible that MO is an example of star formation being reignited in an already star-forming galaxy, this seems unlikely. Although \cite{2006ApJ...647.1040C} could not rule out an underlying old stellar population in MO, an HI cloud downstream from MO was observed, which suggests that unlike Centaurus A the neutral hydrogen cooled out of a warm and clumpy IGM and then collapsed.

While there is observational evidence that in some individual cases intense star formation may have been triggered by R-AGN jets, it is not clear how common this phenomenon is and whether it can be positively stated that the star formation is indeed the result of jet interactions. 
Furthermore, it is not clear if this induced star formation leads to the emergence of pristine, new satellites or whether it is an enhancement in already existing, star-forming satellites. As in the case of star formation quenching, we will approach this question by studying the colors of satellites of R-AGN and a matched control sample of radio quiet galaxies.
In both the study of quenching and the induction of star formation in satellites we will pay special attention to HERGs in which either of these processes may be more pronounced or more common.

The layout of our paper is as follows.  In Section 2,  samples are presented, and in Section 3 we describe our method for selecting the control sample of radio quiet galaxies as well as generating distributions of satellite properties. Our results are presented in Section 4, while in Sections 5 and 6 we discuss the implications of our findings. In this work the cosmological parameters adopted are $\Omega_m\,=\,0.27$, $\Omega_\Lambda\,=\,0.73$, and  $H_0\,=\,71\, \textrm{km}\,\textrm{s}^{-1}\,\textrm{Mpc}^{-1}$.

\section{Construction of Samples}
  
 We have two primary samples: a sample of R-AGN drawn from \cite{2012MNRAS.421.1569B} and a control sample of matching radio quiet galaxies that we have assembled. Each galaxy in the control sample was chosen to have the same type of  environment as its counterpart in the R-AGN sample. Control galaxies were also chosen to approximately match the galaxy mass and star formation histories of the R-AGN by selecting control galaxies that provide the closest match in $r$ magnitude, $u-r$ color, redshift, and axis ratio. We then compare the satellites of R-AGN and radio quiet samples. We use the term `satellite' to refer to nearby galaxies that are physically associated with the hosts in our samples, while `neighbors' refer to candidate satellites, which includes foreground or background objects not associated with the host. This distinction is necessary because candidate satellites are faint and are generally lacking spectroscopic redshifts. We have drawn neighbors for each of these two samples, and to correct for background contamination we have also drawn neighbor-like objects from two sets of offset positions. 
Details of these samples are given below.

\subsection{R-AGN sample}
\label{main}

 \cite{2012MNRAS.421.1569B} have assembled a sample of 18,286 radio galaxies by combining the 7th data release of the SDSS spectroscopic sample (DR7; Abazajian et al. 2009) with the National Radio Astronomy Observatory (NRAO) Very Large Array (VLA) Sky Survey (NVSS; Condon et al. 1998) and the Faint Images of the Radio Sky at Twenty centimeters (FIRST; Becker, White \& Helfand 1995) survey. 
Radio emission may arise from AGN or star formation activity, so \cite{2012MNRAS.421.1569B} used a combination of three methods to separate R-AGN from star-forming galaxies. The first method is a comparison of the 4000 \AA~ break strength to the ratio of radio luminosity per stellar mass. Since R-AGN have enhanced radio luminosity relative to star-forming galaxies, they are thus separable in this plane \citep{2005MNRAS.362...25B}. The second method is the emission-line diagnostic or `BPT' diagram (Baldwin et al. 1981, Kauffmann et al. 2003) which compares the ratio of [O III] 5007 \AA~ and H$\beta$ line fluxes to the ratio of [NII] 6584 \AA~ and H$\alpha$. Only ~30\% of the galaxies in the radio source sample could be classified using the BPT diagram: the remainder lacked detections or limits in at least one line. The final method is a comparison of the H$\alpha$ line luminosity to the radio luminosity. This method operates under the assumption that in star-forming galaxies both the H$\alpha$ line and radio luminosities are a consequence of star formation and thus correlated, while the radio luminosities of R-AGN are offset because their radio luminosities are boosted by AGN activity. Differing classifications are sometimes given by these three methods. In these instances a final classification was chosen based on the properties of the galaxies in each class: see Table A1 of \cite{2012MNRAS.421.1569B}. 
  \begin{figure}[t!]
\includegraphics[width=3.4in]{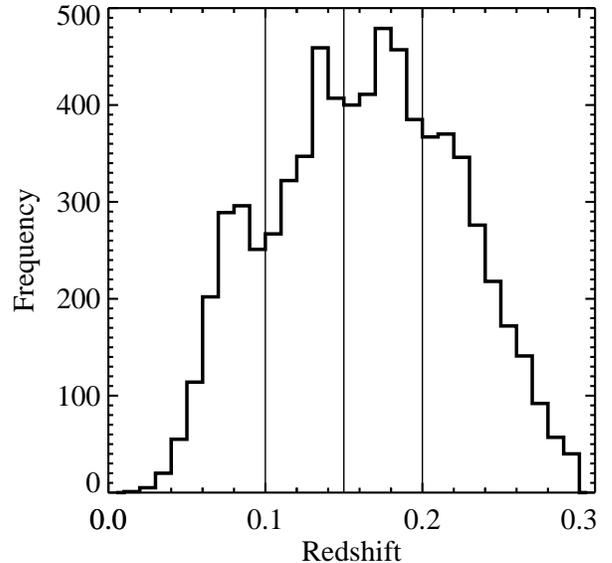}
\caption{ Histogram showing the distribution of R-AGN redshifts. Vertical lines at  $z$ = 0.1, 0.15, and 0.2 indicate divisions used for the redshift ranges.}
\label{reddis}
\end{figure}

  The magnitude range of our sample is 14.5 $<$ \emph{r} $<$ 17.77, matching the limits of the SDSS main galaxy spectroscopic sample \citep{2002AJ....124.1810S}.  Many of the R-AGN in \cite{2012MNRAS.421.1569B} are fainter than this limit and lie at redshifts greater than $z=0.3$. We are not interested in higher redshift R-AGN since we cannot draw control samples for them. We imposed a lower redshift limit at $z=0.04$ to avoid excessive background contamination in the 100 kpc radius used to obtain neighbors. Thus the final redshift range of our sample is 0.04 $<$ \emph{z} $<$ 0.3, and Figure~\ref{reddis} shows this redshift distribution.
 At the lower limit a radius of 100 kpc corresponds to an angular radius of $2\farcm1$. Only 28 R-AGN are removed from the sample as a result of this lower redshift limit. 
  
\begin{figure}[t!]
\includegraphics[width=3.4in]{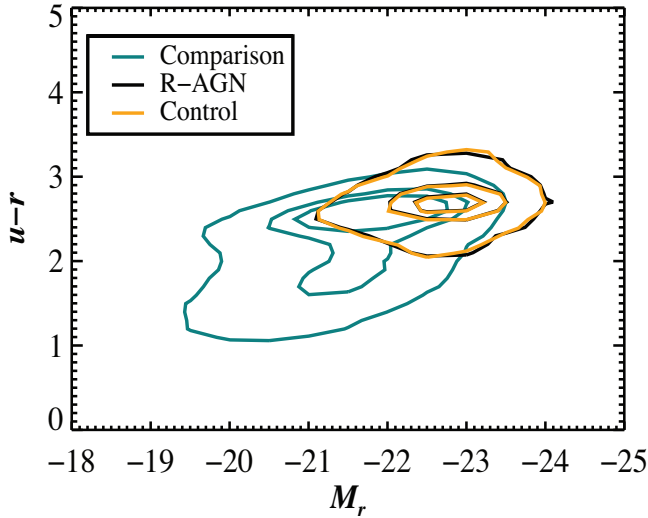}
\caption{Color-magnitude diagram comparing the R-AGN and control samples with SDSS galaxies in the same redshift range. In this figure, $M_r$ and $u-r$ have been $k$-corrected. Contours containing 90\%, 50\%, and 25\% of the samples are shown. R-AGN in the sample tend to be hosted by luminous galaxies on the red sequence. The properties of galaxies in the control sample closely match those of the R-AGN.}
\label{cmd}
\end{figure}

 We used the photometry of the eighth SDSS data release (DR8) \citep{2011ApJS..193...29A} rather than DR7 because photometric redshifts, used to reduce non-satellite contamination in neighbors, tend to be more accurate in DR8 than in DR7 (see Appendix \ref{appa}). Also, in previous data releases the outer regions of large galaxies were oversubtracted, which affected the photometry of these galaxies as well as faint nearby objects \citep{2011ApJS..193...29A}. Improved sky subtraction procedures in DR8 have mostly addressed this issue. Fifty-six objects listed as R-AGN in \cite{2012MNRAS.421.1569B} were either not present or were misidentified in DR8, so our final R-AGN sample consists of 7,220 galaxies.
Figure~\ref{cmd} is a color-magnitude diagram that shows the relation of the R-AGN sample, shown in black, to SDSS galaxies in the same redshift range. Also plotted is the control sample, the selection of which is discussed below. Colors and magnitudes in this figure have been $k$-corrected with the $k$-correction provided by \cite{2012MNRAS.419.1727C}. This figure shows that the R-AGN hosts fall along the red sequence, 
with R-AGN favoring luminous galaxies. 

One of the criteria for producing the control sample is to match the type of environment in which the galaxy is located. Three cluster membership categories were used: field galaxy, cluster member, and BCG. For each of the R-AGN we performed a nearest-neighbor search to the BCGs reported in the galaxy cluster catalog of \cite{2010ApJS..191..254H}, which is based on DR7. The clusters in this catalog are in the redshift range $0.1<z<0.55$, and the list of additional clusters at $z<0.1$, available from the author's web page\footnote{ http://home.fnal.gov/$\sim$jghao/gmbcg\_sdss\_catalog.html}, allow us to assign cluster membership for all R-AGN in the sample. We classified as BCGs all R-AGN whose DR8 coordinates were within 1.5 kpc of the BCG catalog coordinates, and found that our sample of R-AGN contains 949 BCGs or 13\% of the total. Those R-AGN that were between 1.5 kpc and 1.5 Mpc away from the nearest BCG and had a difference in spectroscopic redshift less than  \emph{z} $=$ 0.01 were classified as cluster members. We have chosen 1.5 Mpc as a typical virial radius ($r_{200}$) of clusters. Of our sample, 436 R-AGN or 6\% of the total were classified as cluster members, and the mean distance from cluster members to the BCG of the host cluster is 700 kpc. The remaining 5,835 R-AGN or  81\% of the sample were classified as field galaxies. While R-AGN have a preference for rich environments as demonstrated in previous studies, the majority of radio AGN are nevertheless not in clusters.

Most (86\%) of the R-AGN in the sample have been classified as either HERGs (209) or LERGs (6,005) by \cite{2012MNRAS.421.1569B}.
 The remaining 1,006 were not classified because they lacked emission lines. Table~\ref{tbl1} gives the number of  HERGs, LERGs, and unclassified galaxies in each environment type.
 Interestingly, HERGs are most common among the field R-AGN (3.4\%) and rarer among cluster members (1.1\%). They are very rare among BCGs (0.4\%). This result supports the scenario in which HERGs arise from cold gas accretion, which is more abundant in the field and is in agreement with the findings of \cite{2013MNRAS.430..638S}, who found that LERGs show an increasing rate of incidence with the local density, while HERGs show a decrease, as do optically-selected AGN. We will return to the incidence rate of R-AGNs in different environments in Section~\ref{fraction}.

\begin{deluxetable*}{rcccccccc}
%\centering
\tabletypesize{\scriptsize}
 \tablecolumns{9}
%\tablewidth{ 6.1in}

 \tablecaption{Number of R-AGN per accretion mode and redshift bin for each environment type. \label{tbl1}}
 \tablehead{
    \colhead{Environment}  & \colhead{HERG} & \colhead{LERG} & \colhead{Unclassified}  & \colhead{ $0.04<z<0.1$} & \colhead{ $0.1<z<0.15$}  & \colhead{ $0.15<z<0.2$} & \colhead{ $0.2<z<0.3$}& \colhead{Total}  }%\\    
%    \colhead{Class}&&&&&&&&}

 \startdata
  Field&200&4,845&790&1,115&1,488&1,711&1,521&5,835 \\
%\tableline
Member&5 & 368 & 63&39&129&163&105&436 \\  

BCG &4&792&153&53&185&258&453&949\\
\tableline
Total&209&6,005&1,006&1,207&1,802&2,132&2,079&7,220

\enddata
\tablecomments{ Columns 2 and 3 give the number of R-AGN with the stated accretion mode (from Best \& Heckman 2012) in each type of environment, while column 4 lists those whose accretion mode was not identified. Columns 5-8 present the number of R-AGN in each redshift bin, while column 9 presents the number of R-AGN per environment type. }
\end{deluxetable*}

\subsection{Control sample (non R-AGN)}
For each R-AGN, we have selected a control galaxy which is not an R-AGN (or is below the limits of radio surveys) but has the same cluster membership status and similar overall physical properties as its counterpart R-AGN.
We considered as control candidates those galaxies in DR8 that are within the FIRST footprint ($4^h<RA<20^h$) and therefore subject to classification by \cite{2012MNRAS.421.1569B}. The area of DR8 within the FIRST footprint is identical to that of DR7, so the control candidates are also subject to cluster member classification with the \cite{2010ApJS..191..254H} catalog. This provides a pool of 842,296 control candidates.

Each control galaxy was selected to closely match the $r$ magnitude, $u-r$ color, redshift, and axis ratio of its corresponding R-AGN.  These properties were used for selection because the redshift and $r$ magnitude give the optical luminosity, which together with $u-r$ are a proxy for the stellar mass of the galaxy while the $u-r$ color is a proxy for the star formation history. Radio AGN are predominantly found in spheroidal early-type galaxies (ellipticals) rather than among the flattened early types containing stellar disks (S0s) (e.g., Ekers \& Ekers 1973, V{\'e}ron-Cetty 
\& V{\'e}ron 2001). The two varieties of early type galaxies have similar colors and overlapping absolute magnitude ranges (e.g., Cheng et al. 2011). The inclusion of the axis ratio as a matching criterion helps ensure that the control sample is of the correct morphological type  \citep{2011MNRAS.412..727C}. The selection was done by minimizing the metric,
 \begin{equation}
 \small
 R \,= \,\sqrt{\left(\frac{\Delta z}{0.21}\right)^2+\left(\frac{\Delta r}{2.99}\right)^2+\left(\frac{\Delta( u-r)}{1.96}\right)^2+\left(\frac{\Delta( b/a)}{0.49}\right)^2},
 \end{equation} 
 where the quantities $\Delta z$, $\Delta r$, $\Delta(u-r)$, and $\Delta(b/a)$ are the differences between the R-AGN and control candidate parameters, and the denominators are scaling factors that correspond to the 95 percentile ranges in redshift, $r$ magnitude, $u-r$ color, and axis ratio (b/a).  

The control sample is plotted together with the R-AGN sample in Fig.~\ref{cmd}, and it can be seen that the control sample closely matches the color-magnitude distribution of the R-AGN sample. The means and standard deviations of the differences between the control and R-AGN galaxy properties also demonstrate the success of our control sample selection. The mean difference for $r$ is $\delta r=0.0076$ ($\pm$ 0.065), for redshift the difference is $\delta z=-2.97\times10^{-4}$ ($\pm 4.6\times10^{-3}$), for $u-r$ the difference is $\delta (u-r)=-0.0026$ ($\pm$ 0.05), and for axis ratio the difference is $\delta (b/a)=6.07\times10^{-6}$ ($\pm 1.1\times10^{-2}$). In other words, the standard deviations are roughly 2\% of the 95 percentile range used for the scaling factors in the metric. Some of the control galaxies are best matches to more than one of the R-AGN in the sample: this occurs for only 11\% of the control galaxies. 

\begin{verbatim}

\end{verbatim}
\subsection{Satellites of R-AGN and control samples}
\label{radius}
%\subsubsection{Radius for neighbor search}

In this study, we have examined all satellites within a circular region centered on the host galaxies in our samples regardless of their position angle with respect to the jet. For the study of feedback on the satellite population it may appear better to focus only on satellites lying close to the jet. However, jet directions are not available for the majority of the sample. Furthermore, the satellite PKS2250-41 \citep{2008MNRAS.386.1797I} shows evidence of jet-induced star formation even though it is not currently in the jet path, justifying our approach. 
The radius of the search circle was determined by examining the projected lengths of the 782 R-AGN jets in the list provided by \cite{2002A&A...381..757L}.
Of the R-AGN in this list, 90\% have jets shorter than 100 kpc. A search radius greater than 100 kpc would allow us to include the end points of a few more R-AGN at the cost of a large increase in unassociated neighbors. We restricted our analysis to satellites that lie within 100 kpc (in projection) of the galaxies in our samples. An inner boundary at 5 kpc was also used to avoid spurious sources which may arise as a result of shredding of the host galaxy by the SDSS pipeline. 
Neighbors include objects classified as galaxies by the DR8 pipeline that are brighter than $r\,=\,22$, close to the SDSS magnitude limit of $r\,=\,22.2$. We make no requirement that the magnitudes of neighbors be fainter than that of the  R-AGN or control galaxy, but we find that only in a few instances are neighbors more luminous than their target galaxies.

We derive properties of the satellite population by subtracting the distribution of sources selected in the same way as neighbors, but in parts of the sky that are offset (two degrees in RA) from the host (R-AGN or its control). To make this procedure more robust, we first remove from candidate neighbors those with significantly discrepant photometric redshifts. Considering the large uncertainties of available photometric redshifts (Appendix A) we apply a very conservative cut such that the candidate neighbors are removed if their photometric redshift is more than 0.2 greater than the spectroscopic redshift of the host.

\section{Methods}
\label{Methods}

In this study we examine the co-added distribution of satellite properties for R-AGN, and, separately, for their radio quiet counterparts, which span a redshift range from 0.04 to 0.3. As a result, at the low redshift limit we probe galaxies down to $M_r$ = $-14.3$, which shifts to $M_r$
 = $-18.8$ at the upper redshift limit. To account for the resulting Malmquist bias we first construct various satellite distribution functions in smaller redshift ranges, in which the bias is greatly reduced. After we confirm that different redshift slices have similar distributions of satellite properties, we average them.
The four redshift ranges are 0.04 $<\,z\,<$ 0.1, 0.1 $<\,z\,<$ 0.15, 0.15 $<\,z\,<$ 0.2, and 0.2 $<\,z\,<$ 0.3. The distribution of R-AGN redshifts as well as the redshift range divisions are shown in Fig.~\ref{reddis} while Table~\ref{tbl1} gives the number of R-AGN per environment type in each redshift range. For a given redshift range, the number of neighbors per bin of a distribution are given by,
\begin{equation}
N_R\,=\,\textrm{Number of R-AGN neighbors},
\end{equation} 
\begin{equation}
N_c\,=\,\textrm{Number of control neighbors},
\end{equation} 
\begin{equation}
B_R\,=\,\textrm{Number of R-AGN offset objects},
\end{equation} 
\begin{equation}
B_c\,=\,\textrm{Number of control offset objects}.
\end{equation} 

Table~\ref{tbl2} gives the number of neighbors and offset objects per redshift bin in each environment. This table shows that roughly half of neighbors are background objects.
The number of satellites per host around R-AGN is given by subtracting the offsets from the number of R-AGN satellites and dividing by the number of primary galaxies in that redshift range, 
\begin{equation}
N_{sat,i}\,=\,\frac{N_R-B_R}{N_{host}}.
\end{equation} 
To get the error in the number of satellites per host, we use, assuming normal approximation of counting errors:
\begin{equation}
\sigma_{sat,i}\,=\,\frac{\sqrt{N_R+B_R}}{N_{host}}.
\end{equation} 
The same method is used to calculate the number of satellites per control galaxy. However, since in 11\% of cases the same control is chosen for more than one R-AGN the errors calculated in the above way, which assumes independent measurements, will be underestimated for control galaxies by 5\%. This is a small effect and we do not correct for it. This procedure gives the nearly volume-complete distributions of R-AGN and control satellites separately for each of the four redshift ranges. We then combined the four distributions with weighted averaging, and the resulting distribution for the R-AGN sample is given by 

\begin{equation}
D_{final,R}\,=\,\frac{\sum_{i=1}^{4}D_{R,i}/\sigma^2_{R,i}}{\sum_{i=1}^{4}1/\sigma_{R,i}^2},
\end{equation}

and the error is given by
\begin{equation}
\sigma_{final,R}\,=\,\frac{1}{\sum_{i=1}^{4}1/\sigma_{R,i}^2}. 
\end{equation} 
\begin{verbatim}

\end{verbatim}
The final distribution for the control sample is computed in the same manner. 

\begin{deluxetable}{rcccccccc}[h]
%\centering
\tabletypesize{\scriptsize}
 \tablecolumns{6}
%\tablewidth{ 6.1in}

 \tablecaption{Number of neighbors and offsets per redshift bin brighter than $M_r=-17$. Neighbors with discrepant redshifts are not included in these counts.\label{tbl2}}
 \tablehead{
   \colhead{Environment} & \colhead{Number}   & \colhead{ Bin 1} & \colhead{ Bin 2}  & \colhead{ Bin 3} & \colhead{ Bin 4}  }%\\    
%    \colhead{Class}&&&&&&&&}

 \startdata
&  $N_R$&10,023&7,899&7,337&5,815\\
Field & $N_C$&9,203&6,001&6,789&4,825\\
  &$B_R$&5,820&3,935&3,386&2,485\\
  &$B_C$&5,906&4,108&3,525&2,490\\
  \tableline
 &  $N_R$&458&867&915&543\\
Cluster& $N_C$&437&784&692&454\\
 Member &$B_R$&183&321&308&191\\
  &$B_C$&247&367&298&164\\
  \tableline
  &  $N_R$&707&1,690&1,732&2,605\\
BCG & $N_C$&658&1,503&1,577&2,330\\
  &$B_R$&246&535&528&712\\
  &$B_C$&203&549&487&709

\enddata
\tablecomments{ These redshift bins are the same as those in Table~\ref{tbl1}.}
\end{deluxetable}

\section{Results}
\label{results}

To investigate R-AGN triggering and the effects of R-AGN jets on satellite galaxies, we compare the populations of R-AGN and control satellites. This is done by calculating and comparing distributions of satellites of R-AGN and control galaxies as functions of projected distance from the host, satellite luminosity, and satellite color. We have also examined the ratios of distributions of R-AGN and control satellites.

\subsection{Distribution of projected distances of satellites from the host}

\begin{figure*}[t!]
\centering
\includegraphics[width=4.5in]{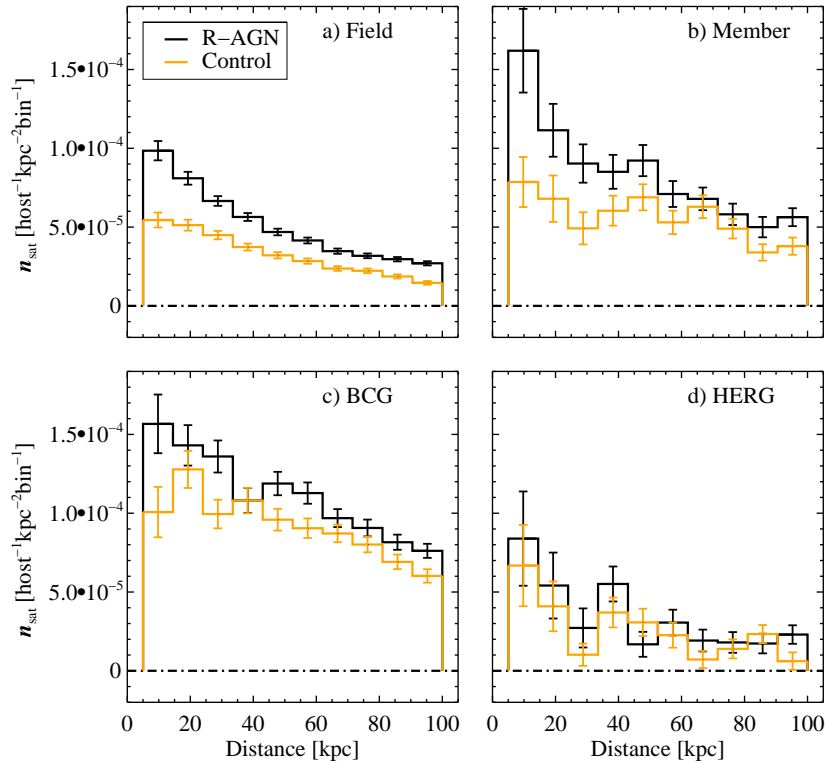}
\caption{Histograms of the distance distributions of satellites for which $M_r < -19$, our completeness limit in the highest redshift bin. The process of filtering by photometric redshift, subtracting the offset, and combining redshift ranges with weighted averaging was used to create these distance distributions.  Distributions in black are for R-AGN satellites, while those in orange are for the satellites of control galaxies. Panels a, b, \& c show the results for field galaxies, cluster members, and BCGs, respectively, while panel d shows the results for HERGs. These plots show that R-AGN in all environments have more satellites within 100 kpc than radio-quiet galaxies.}
\label{distance}
\end{figure*}

Figure~\ref{distance} shows distributions of projected radial distances of satellites from their hosts. Field galaxies are in panel a, cluster members in panel b, BCGs in panel c, and HERGs, regardless of environment, in panel d. At all redshifts, our satellite samples are complete for satellites brighter than $M_r \,=\, -19$, so we adopt this absolute magnitude cut for the radial distance distributions. It should be noted that 90\% of the HERGs are field galaxies. As expected, BCGs have more satellites than either field galaxies or non-BCG cluster members, regardless of whether they host an R-AGN or not. The satellite populations of field galaxies and non-BCG cluster members are similar for a given type of host. However, what is remarkable is that the mean number of satellites differs between R-AGN and radio-quiet galaxies, and it does so in all environments.
Furthermore, this excess of satellites around R-AGN is present over the entire 100 kpc projected radius that we explore. The situation is ambiguous for the case of HERGs, where no obvious excess is visible.
What we can confidently say is that the excess seen in panels a, b, and c reflects differences in the satellite populations of LERGs and not HERGs.

\subsection{Absolute magnitude distribution of satellites}

\begin{figure*}[t!]
\centering
\includegraphics[width=4.5in]{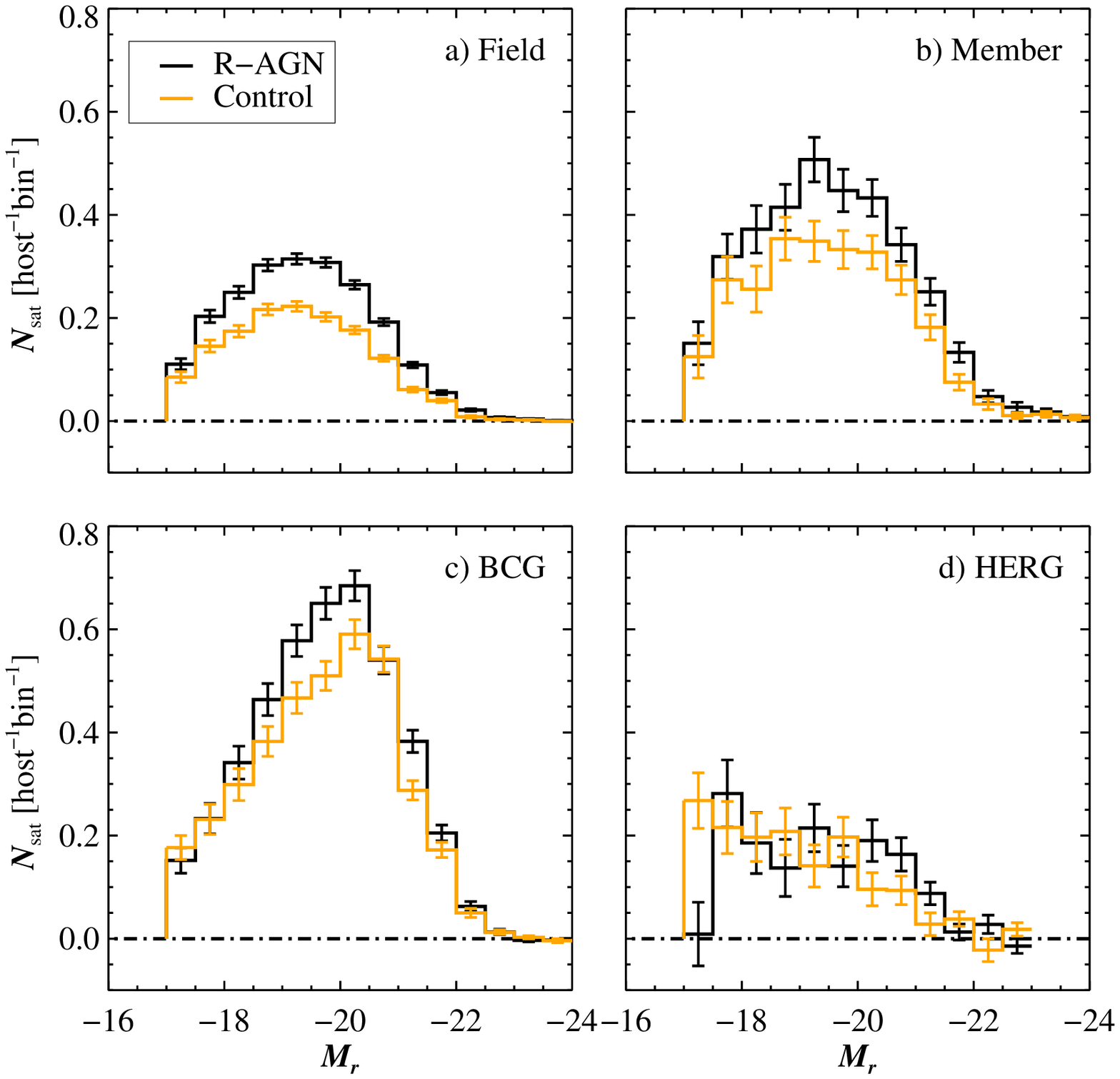}
\caption{Histograms of the absolute magnitude distributions of satellites for which $M_r\leq-17$, our completeness limit in the lowest redshift bin. Panels a-d are as in figure~\ref{distance}. Distributions of R-AGN satellites are shown in black, with control satellites in orange. As with Figure~\ref{distance}, these plots show that R-AGN in all environments have more nearby satellites than radio-quiet galaxies, and that this excess is not restricted to satellites of some luminosity.}
\label{abmag}
\end{figure*}

We construct satellite luminosity distributions to further characterize the differences in the satellite populations of R-AGN versus the control sample. Absolute magnitudes were computed using the spectroscopic redshifts of the host galaxies and have been $k$ corrected using the corrections provided by \cite{2012MNRAS.419.1727C}. 
The resulting $M_r$ distributions are shown in  Fig.~\ref{abmag} with the results for the three environments in panels a, b, \& c, and HERG satellites in panel d. 

Unlike the luminosity function of field galaxies, the $M_r$ distributions of satellites in Fig.~\ref{abmag} have a peak, which is to be expected for a flux-limited sample \citep{2006ApJ...647...86C}. 
The distribution of BCG satellites peaks at a brighter magnitude than field galaxies or cluster members. Since BCGs are generally massive and luminous, their satellites tend to be more luminous as well. The distributions of the satellites of R-AGN and control galaxies are quite similar among themselves, for all environments, with peaks at similar positions.  
In terms of the average number of satellites per host that are brighter than $M_r=-17$, we find that field R-AGN have 2.14 ($\pm0.03$) satellites per host, and field control galaxies have 1.46 ($\pm0.03$). For cluster members, R-AGN have 3.47 ($\pm0.12$) and control galaxies have 2.61 ($\pm0.11$) satellites per host, while R-AGN that are BCGs have 4.30 ($\pm0.09$) and control BCGs have 3.72 ($\pm0.08$) satellites per host. We find that HERGs have 1.43 ($\pm0.15$) satellites per host on average, while the control galaxies of HERGs have 1.48 ($\pm0.13$). This formally confirms that the excess is very significant in all environments. However, no excess is formally found for HERGs. 

\subsection{Satellite colors}
\begin{figure*}[t!]
 \centering
\includegraphics[width=4.5in]{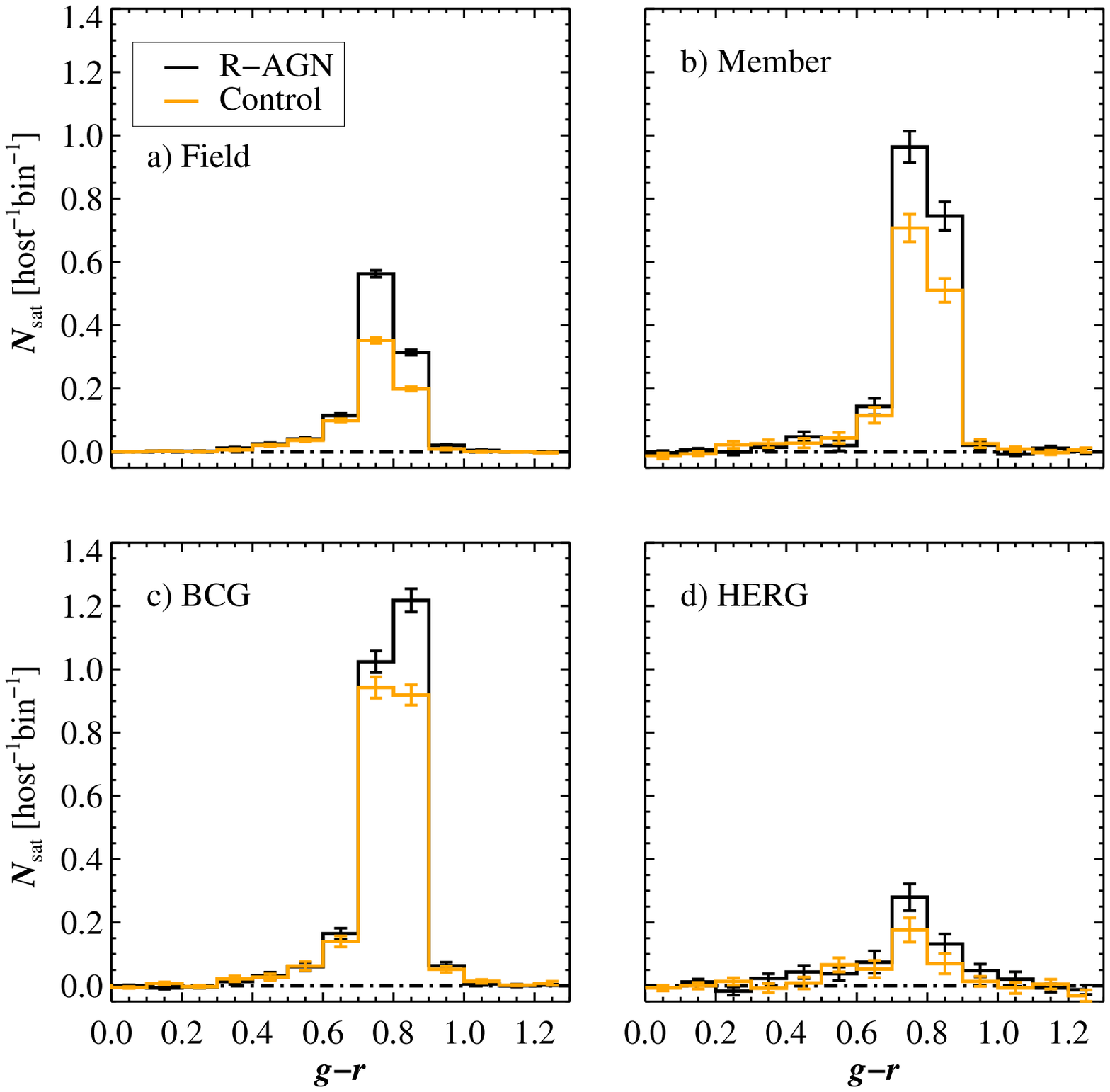}
\caption{ Distributions of $k$-corrected $g-r$ colors for R-AGN and control satellites. These plots are for satellites for which $\sqrt{\sigma_g^2+\sigma_r^2}< 0.1$ and $M_r< -19$. Panels a-d are as in figure~\ref{distance}. The majority of satellites are red. }
\label{color}
\end{figure*}

We compared the $g-r$ color distributions of R-AGN and control satellites to provide context for the examination of the feedback effects of R-AGN jets on nearby satellites (Section~\ref{feedback}). Although $u$ is more sensitive to recent star formation, we chose to use the $g-r$ colors for satellites since many of them have poor $u$-band photometry.
We again used an absolute magnitude cut at $M_r\,=\, -19$ to probe the same part of the satellite population at all redshifts. The $k$-correction of \cite{2012MNRAS.419.1727C} was also applied to the $g-r$ colors, and the analysis was restricted to satellites for which $\sigma (g-r) < 0.1$ mag ($\sim$90\% of total). Histograms of these distributions are shown in Fig.~\ref{color}.

In general the distributions of both the satellites of R-AGN and control galaxies peak near $g-r=0.8$, indicating that satellites in general are red irrespective of R-AGN. 
In other words, the satellite population around these massive galaxies is mostly quenched. Whether R-AGN feedback plays any active role in this quenching cannot be well established from these distributions and is deferred to Section~\ref{feedback}.
The distribution for HERG satellites is noisy but is consistent with no excess satellites around HERGs, as seen in panel d of Fig.~\ref{distance} where the same $M_r$ cut was applied.

\subsection{Characterization of the excess satellite population}
\begin{figure*}[t!]
\centering
\includegraphics[width=4.5in]{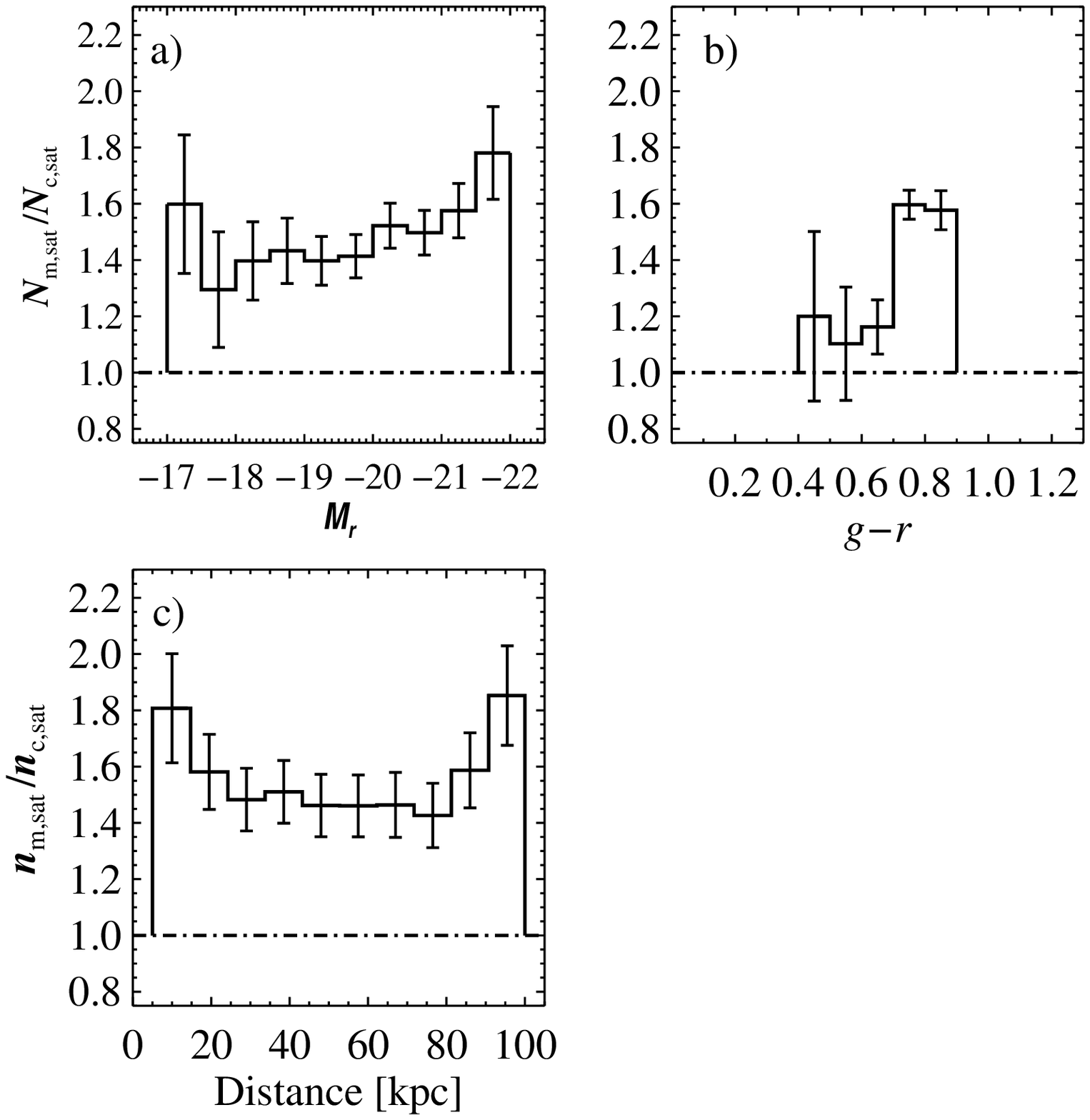}
\caption{Ratios of the distributions of satellites for field R-AGN and their control, radio-quiet galaxies. Panel a shows the ratio of luminosity distributions, panel b the color distributions, and panel c the distance distributions. The ratio is typically $\sim$1.5 with a possible weak dependence on brightness and color. See text for more details.}
\label{ratio}
\end{figure*}

From the number histograms alone it is difficult to see if the population of excess R-AGN satellites has a preferred separation from the host, a preferred luminosity, or color. This is because the absolute excess will appear stronger where the distribution is higher.
We have therefore examined the ratio of the distributions of R-AGN and control satellites.  Ratio plots of the $M_r$, color, and distance distributions for the field sample are shown in Fig.~\ref{ratio}, with error bars computed via simple error propagation. Individual bins with errors greater than 50\% are not shown.  We have only shown plots for the field sample because it contains the most R-AGN: similar results are seen for the member and BCG samples.  

All three distributions in Fig.~\ref{ratio} are roughly flat at 1.5, which indicates that  field R-AGN have $\sim$50\% more satellites than control galaxies. In other words, the satellite populations of R-AGN are scaled up versions of the satellite population around radio-quiet galaxies. We may be seeing a slight preference of the excess population for brighter and redder satellites. The former is not surprising as any possible influence on AGN activity would presumably be more likely for more massive satellites. The latter can be explained if some of the blue satellites are preferentially further from a galaxy in real distance than their projected distance would suggest and consequently do not contribute to excess population.

%We may be seeing a slight preference of the excess population for brighter and redder satellites. The former is not surprising as any possible influence on AGN activity would presumably be more likely for more massive satellites. The latter can be explained if some of the blue satellites are preferentially further from a galaxy in real distance than their projected distance would suggest and consequently do not contribute to the excess population.

\section{Discussion}

\subsection{R-AGN fraction}
\label{fraction}

\begin{figure}[h!]
\includegraphics[width=3.25in]{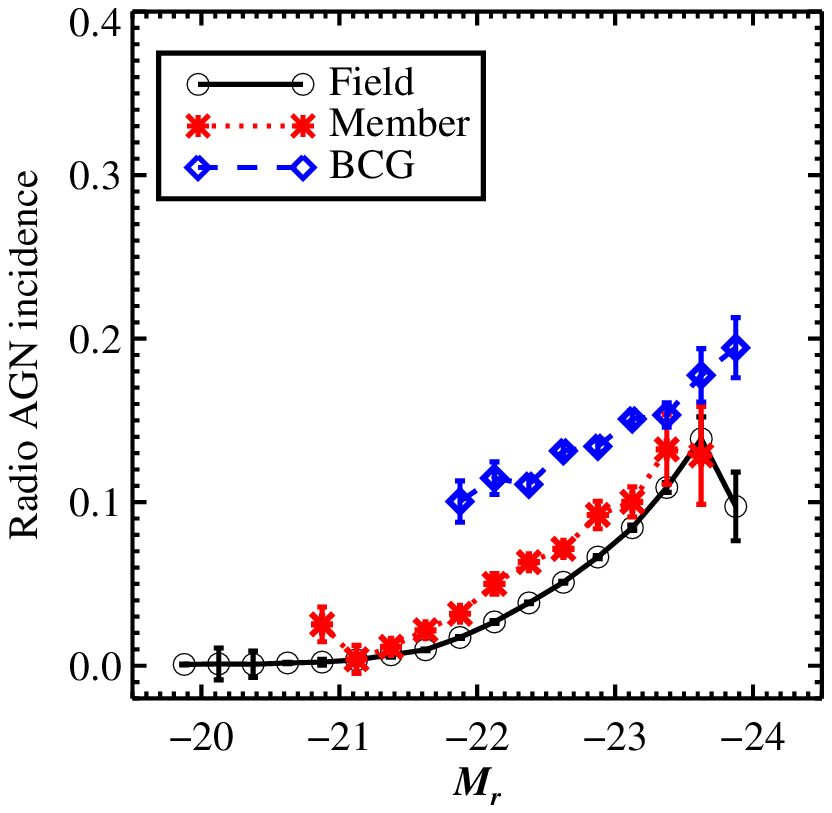}
\caption{Fraction of galaxies that are R-AGN, for different environments. The fraction is calculated as the median of individual incidence estimates. These estimates are the ratio of galaxies that are R-AGN among galaxies that have similar properties (redshift, magnitude, color, axis ratio), i.e. galaxies eligible to be R-AGN. The R-AGN fraction is higher in BCG's, with cluster members having a slightly higher incidence than field galaxies. }
\label{ragnfrac}
\end{figure}

Previous studies (e.g. Best et al. 2005b, Van Velzen et al. 2012) have shown that the fraction of galaxies that are R-AGN (the R-AGN incidence rate) is an increasing function of the stellar mass of the host. We revisit these results by including two new aspects in the analysis. First, we obtain incidence rates separately for R-AGN in each type of environment (BCGs, cluster members, and field). Second, we consider the incidence rate to be the R-AGN fraction in the {\it eligible} galaxy population, not among all galaxies in a given mass (i.e., luminosity) bin. The idea is that one is primarily interested in R-AGN incidence among the galaxies that could (and perhaps did or will) host an R-AGN, so we define the eligible population to occupy the same part of the parameter space in redshift, magnitude, color, and axis ratio as galaxies that are {\it current} R-AGN hosts. Each R-AGN in the sample defines a position in the parameter space around which we determine the R-AGN incidence. We take the vicinity to be $R<0.2$ (Eq. 1) and count the R-AGN used to probe the parameter space as 0.5, which leads to less bias than either not counting it or giving it a full count. These fractions are then binned by $M_r$, and the median of each bin is shown in Fig.~\ref{ragnfrac}. Bins with fewer than five R-AGN were omitted.

Figure~\ref{ragnfrac} shows that while the fraction of eligible galaxies that currently host an R-AGN is generally low, with an overall average of 6\%, the incidence increases from close to zero to a high near $\sim$20\% for the most luminous galaxies. The incidence is generally higher in BCGs, as might be expected, while the incidence in cluster members is only slightly greater than that of field galaxies. This suggests that fueling is primarily a \textit{locally} determined process and that being in a more massive cluster halo confers modest additional benefits, unless a galaxy sits in the center of the cluster halo, as is the case for BCGs.

\cite{2005MNRAS.362...25B} show a similar relation for R-AGN incidence but as a function of stellar mass in panel a of their Fig. 2. The mean NVSS radio luminosity of our sample is $2\times10^{24}$ W Hz$^{-1}$ at 1.4 GHz, so our results are best compared to the middle curve of panel a. The R-AGN incidence of \cite{2005MNRAS.362...25B} is expressed as a fraction of all galaxies from SDSS DR2 without regard
\begin{verbatim}


\end{verbatim}
for the galaxies' likelihood of hosting R-AGN, while we have only considered galaxies from the same type of environment that are likely to host R-AGN. Even so, considering that most R-AGN are field galaxies and the majority of massive galaxies are eligible as R-AGN, the results for the field galaxies and cluster members in our sample follow a trend similar to that of \cite{2005MNRAS.362...25B} for $10^{24}$ W Hz$^{-1}$ at 1.4 GHz.  

\cite{2007MNRAS.379..894B} found that at a stellar mass of $\sim5\times10^{11}M_\sun$ the R-AGN incidence in BCGs is only slightly greater than that of other galaxies, although for masses lower than $\sim\times10^{11}M_\sun$ the R-AGN incidence in BCGs is over an order of magnitude greater than that of other galaxies. This is in agreement with what we find in Fig~\ref{ragnfrac}. \cite{2007MNRAS.379..894B} also found that except for group and cluster galaxies within $0.2r_{200}$ of the center of the system, the R-AGN incidence among non-BCG group and cluster galaxies is similar to that of field galaxies. This was based on a comparison of the R-AGN sample defined by \cite{2005MNRAS.362....9B} to all galaxies in DR4 within the same redshift range as the \cite{2005MNRAS.362....9B} R-AGN sample, regardless of these galaxies' likelihood of hosting an R-AGN. These results are in agreement with our finding that field and cluster member R-AGN have similar incidence rates.

\subsection{The role of satellite populations in R-AGN triggering}

We have found that in all environments R-AGN have more nearby satellites than radio-quiet galaxies, but our results are not a simple confirmation of what other studies have found regarding the R-AGN clustering. \cite{2008MNRAS.391.1674W}, \cite{2009MNRAS.393..377M}, \cite{2010MNRAS.407.1078D}, and \cite{2008MNRAS.384..953K} find that R-AGN are more clustered than optical AGN or radio-quiet galaxies. The control samples of \cite{2009MNRAS.393..377M}, \cite{2010MNRAS.407.1078D}, and \cite{2008MNRAS.384..953K} were even chosen to match the stellar masses of the R-AGN. However, the results of these studies can be interpreted simply to mean that R-AGN prefer denser, cluster environments or, judging by the results in Section~\ref{fraction} that show that R-AGN incidence among eligible cluster members and field galaxies is the same, that the type of galaxies that typically host R-AGN (massive ellipticals) are more clustered, a well known result \citep{1980ApJ...236..351D}. By including cluster membership in the selection criteria of control galaxies we have added crucial new information: the satellite population of R-AGN is richer, whether the R-AGN is a field galaxy, cluster member, or the very BCG.

It is now widely believed that LERGs are triggered by the cooling of small amounts of gas from the hot halos in which they reside \citep{2012MNRAS.421.1569B}. Since $\sim$80\% of the R-AGN in our sample are LERGs, our finding that R-AGN in all environments have an excess of satellites suggests that the availability of hot halo gas is not the only prerequisite for triggering activity in LERGs. For example, the dark matter halo masses of the BCGs in our R-AGN and control samples are similar, which would suggest that they have similar quantities of hot halo gas available. The difference is that BCGs that host R-AGN have more nearby satellites, which therefore probably play some role in triggering. This argument applies to field galaxies and cluster members as well, since the matched pairs of R-AGN and control galaxies have similar halo masses and therefore presumably similar amounts of hot halo gas. Again, the difference between R-AGN and radio-quiet galaxies in these environments is the number of nearby satellites. Results in Section~\ref{results} suggest that the triggering is facilitated by satellites with a wide range of masses and at various projected distances from the host.

\subsection{Dark matter halo bias}
While it is tempting to interpret the excess of satellites as related to the presence of R-AGN, we must explore the possibility that even if the control sample is matched in stellar mass, it may be systematically different in halo mass. If the abundance of satellites follows the halo mass rather than the stellar mass, the excess of satellites may result from a mismatch in halo masses. 
We have used the dark matter halo catalog of \cite{2007ApJ...671..153Y} to investigate this possibility. This catalog presents two dark matter halo masses for each galaxy: one calculated based on the ranking of group characteristic luminosity, and the other based on the ranking of group stellar mass: we used those based on the group stellar mass. This catalog uses the fourth SDSS data release (DR4) and extends out to $z=0.2$, so only about 1,400 matched pairs of R-AGN and control galaxies were found. The mean of the log of R-AGN masses is 0.1 dex greater than that of control galaxies. Is this modest systematic offset in halo masses sufficient to explain a factor of 1.5 difference in satellite population?

\begin{figure}[t!]
\centering
\includegraphics[width=3.25in]{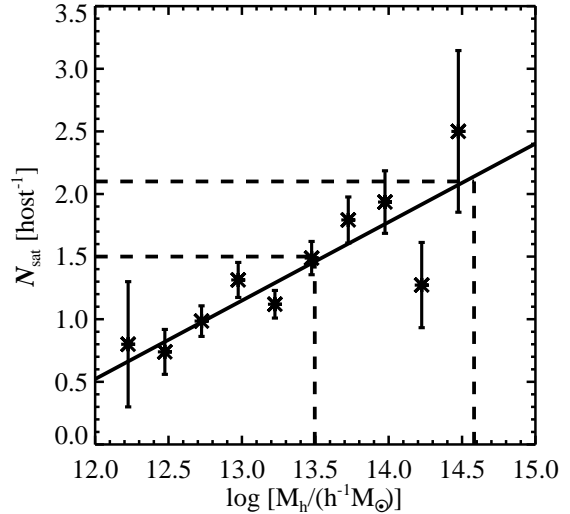}
\caption{ Distribution of satellite counts as a function of halo mass for the field sample of control galaxies in the redshift range $0.15<z<0.2$. The solid line is the least-squares fit to the distribution, and the dashed horizontal lines show the average number of satellites for R-AGN (2.1) and control galaxies (1.5). For the excess R-AGN satellites to be a result of mismatched halo masses, a difference of 0.96 dex (dotted vertical lines) between R-AGN and controls is necessary. Since the actual difference is only 0.1 dex, R-AGN must genuinely have more satellites within 100 kpc than radio-quiet galaxies. }
\label{halo}
\end{figure}

Figure~\ref{halo} shows the number of satellites as a function of halo mass for the field control sample in the redshift range $0.15<z<0.2$. There is a trend that galaxies in more massive halos indeed have more satellites. However, this rise is not so steep. In order to bring the number of satellites from 1.5 (the number observed on average around control sample galaxies) to 2.1 (the number for R-AGN hosts) the mismatch in halo masses would have to be 0.96 dex (a factor of 9) while in reality it is only 0.1 dex (25\%). In other words, the excess of R-AGN satellites cannot simply be a result of R-AGN having dark matter halos that are systematically more massive than those of control galaxies.

\subsection{The effects of R-AGN feedback on the satellite population}
\label{feedback}
In addition to the triggering of R-AGN, we also wish to examine the satellite population with the goal of understanding the effects of R-AGN interactions (presumably that of the jet) with satellites. In the discussion that follows, we use color distributions of satellites to examine two scenarios. In the first, R-AGN jets frequently quench star formation in their satellites, while in the second interactions with R-AGN jets lead to induced star formation.

\begin{figure*}[t!]
\centering
\includegraphics[width=6.1in]{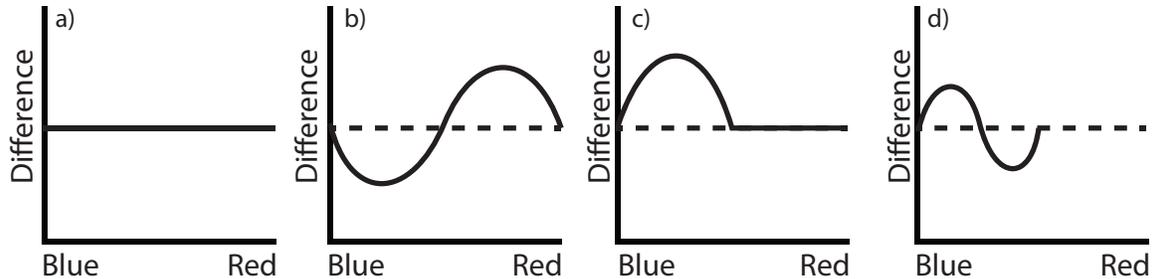}
\caption{ Schematic plots illustrating the effects of R-AGN feedback on the difference color distribution between R-AGN and control galaxies, assuming no net difference in the number of satellites. Panel a shows the case where neither quenching nor induced star formation as a result of jet interactions are common in satellites. Panel b shows the result of star formation being quenched in many R-AGN satellites as a result of jet interactions, leading to a deficit of bluer satellites that is matched by an excess of red ones, while panel c shows the result of radio jets inducing star formation in gas clouds that would not otherwise form stars, producing a net excess of blue satellites. Panel d shows the result of radio jets enhancing star formation in satellites that are already actively forming stars.}
\label{same}
\end{figure*}

We study the effects of R-AGN feedback by comparing the difference between $g-r$ color distributions of R-AGN and control satellites. 
Let us first examine how feedback would modify the difference in color distribution between R-AGN and control galaxies in the absence of an overall excess of satellites around R-AGN. 
If neither quenching nor induced star formation are present, the difference would be zero, as in panel a of Fig.~\ref{same}.  
If interactions commonly lead to quenching, there would be a deficit of blue R-AGN satellites accompanied by an increase in the number of red R-AGN satellites relative to the control distribution, as shown in panel b. 

\begin{figure*}[t!]
\centering
\includegraphics[width=6.1in]{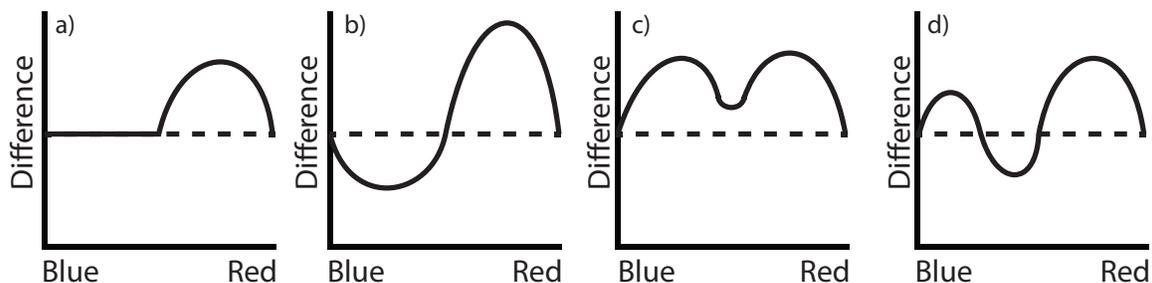}
\caption{  Schematic plots illustrating the effects of R-AGN feedback on the difference color distribution between R-AGN and control galaxies, for scenario in which R-AGN have more red satellites than control galaxies. Panel a shows the case where neither quenching nor induced star formation as a result of jet interactions are common in satellites. Panel b shows the result of star formation being quenched in many R-AGN satellites as a result of jet interactions, leading to a deficit of bluer satellites that is matched by an excess of red ones, while panel c shows the result of radio jets inducing star formation in gas clouds that would not otherwise form stars, producing a net excess of blue satellites. Panel d shows the result of radio jets enhancing star formation in satellites that are already actively forming stars.}
\label{radiomore}
\end{figure*}

Next, we consider two scenarios of induced star formation. In the first, radio jets induce star formation in gas clouds that would not otherwise be detected as galaxies, i.e. induced star formation forms new galaxies. 
%This scenario also applies to faint satellites that experience an increase in luminosity as a result of induced star formation and thus become detectable. 
Induced star formation would cause R-AGN to have an excess of blue satellites, as shown in panel c. Faint satellite galaxies that increase in luminosity as a result of induced star formation would also lead to an excess of blue satellites. 
In the other scenario, interactions with R-AGN jets may enhance the SFR in satellites that already form stars, making them bluer (panel d). In this scenario there is no net excess, but some moderately blue galaxies become bluer, causing a deficit at moderate value colors and an excess at very blue colors. 

However, from Section~\ref{results} we know that there is an intrinsic excess of satellites around R-AGN, so the scenarios described above need to be modified to take that into account, as shown in Figure~\ref{radiomore}. As we have seen, this excess is primarily red simply because most satellites are red, so we obtain modified scenarios by adding a red bump to the scenarios described above. Now these modified scenarios can be compared to the observations.

\begin{figure*}[t!]
\centering
\includegraphics[width=4.5in]{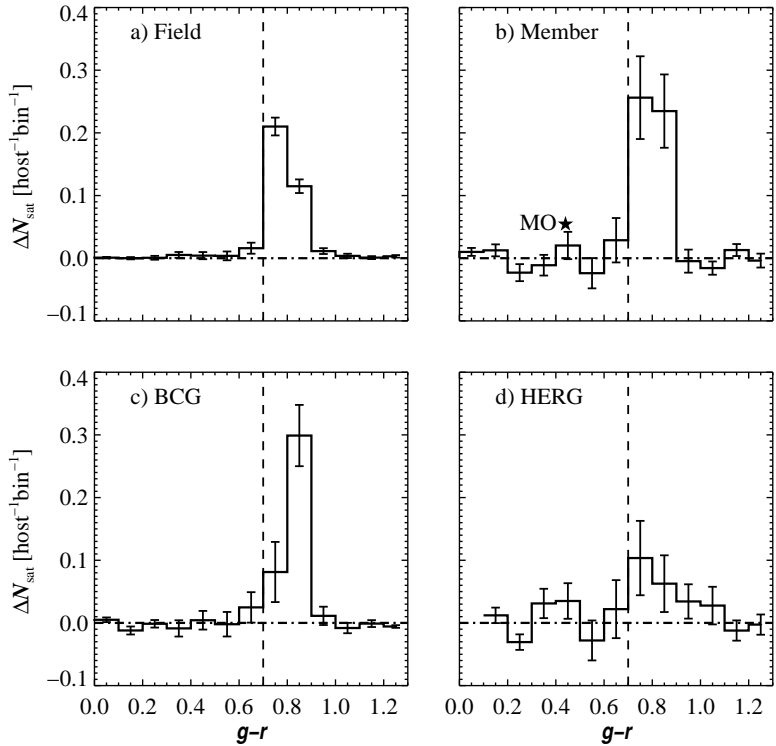}
\caption{Plots of the difference between the color distributions of R-AGN and control satellites: panels a-c are for the three environment types and panel d is for HERGs. The dotted vertical line at $g-r=0.7$ marks the split between blue and red satellites. Panels a-c resemble panel a of Fig~\ref{radiomore}, indicating that R-AGN neither quench nor induce star formation in their satellites. The point labelled MO in panel b shows what would be 5\% occurrence of satellites with the color of Minkowski's Object. The actual value is much closer to zero, which suggests that R-AGN rarely induce star formation in external gas clouds. Quenching as a result of jet interactions is also uncommon.}
\label{minus}
\end{figure*}

Figure~\ref{minus} shows the difference plots for the three environment types as well as the HERG sample. The vertical dotted line at $g-r=0.7$ indicates the split between blue and red satellites such that these plots can be easily compared to Fig.~\ref{radiomore}.  Panels a$-$c of Fig.~\ref{minus} resemble panel a of Fig.~\ref{radiomore}, with no blue deficit accompanying the red bump. This indicates that in any environment, LERGs play no role in quenching star formation in satellites. 

\cite{2011MNRAS.413.2815S} examined whether R-AGN quench star formation in satellite galaxies that lie in the projected path of the radio jets. These authors manually examined the satellites of a small sample of 21 FR1 and 72 FR2 sources and classified them, based on their spatial relation to the radio jets as lying inside or outside the jet path. It was found that the $u-r$ color distribution of satellites in the path of FR1 radio jets is similar to that of satellites outside the path while satellites in the jet path of FR2s are redder than satellites outside the path. \cite{2011MNRAS.413.2815S} therefore concluded that only the high-power FR2 galaxies quench star formation in satellites. 
If we assume that most of the HERGs in our sample present FR2 morphology, then we may compare our findings for HERG satellites to the results of \cite{2011MNRAS.413.2815S}. Panel d of Fig.~\ref{minus} shows that, like R-AGN in all environments, HERGs have an excess of red satellites. This excess is not accompanied by a deficit of blue R-AGN satellites, as the sum of the bins bluer than $g-r=0.7$ is 0.04 ($\pm0.07$). Our results indicate that high accretion R-AGN, like low accretion ones, at best only rarely quench star formation in nearby satellites.
This non-detection of quenching is in apparent disagreement with the results of \cite{2011MNRAS.413.2815S}, but note that we have considered quenching on entire satellite populations, while \cite{2011MNRAS.413.2815S}  only investigated satellites in the path of the jet.

If R-AGN frequently induced star formation in their satellites, we would expect an excess of blue satellites between $0\lesssim g-r\lesssim 0.6$ in Fig.~\ref{minus}. Such an excess is not seen for R-AGN in any environment, nor for HERGs. The field sample places a limit on the incidence of star formation induction to no more than 1\%. This provides evidence that star formation is rarely induced by interactions with R-AGN jets. We have included in panel b of Fig.~\ref{minus} a point showing what would be 5\% occurrence of satellites with the color of Minkowski's Object. The actual value is consistent with zero, which indicates that R-AGN rarely induce star formation in external gas clouds, nor do they lead to any significant enhancement in existing gas-rich satellites.

\section{Summary}
We have studied R-AGN triggering by comparing the satellite populations of R-AGN and a control sample of radio-quiet galaxies, which was matched to the R-AGN in luminosity, color, redshift, axis ratio, and cluster membership. We also compared these two satellite populations to search for evidence of quenching in satellites as a result of jet interactions as well as jet-induced star formation in satellites. We separately analyzed HERGs to determine whether either the triggering or the feedback behaves differently for powerful radio AGN.

\noindent Our conclusions can be summarized as follows:

\begin{enumerate}

\item The incidence rate of R-AGN, which we define as the fraction of R-AGN among the galaxies whose properties are such that they could host an R-AGN and are found in a similar environment, is a strong function of luminosity, as found in previous studies that defined the incidence rate in a less restrictive way (e.g., with respect to all galaxies in a mass bin).

\item We find that the incidence rate of field R-AGN and cluster member R-AGN is comparable (4\% for field, 6\% for members on average), while that of BCGs is significantly higher at the same luminosity (14\% on average). The relative similarity of the incidence rates of field and non-BCG cluster members suggests that some previous studies that find that R-AGN are more clustered on cluster scales (~1 Mpc) either derive that signal from BCGs, which are indeed more likely to be R-AGN, or from the fact that the type of galaxies that host R-AGN (massive ellipticals) are more likely to be found in clusters. The incidence of R-AGN is consequently primarily dependent on the availability of fuel (hot gas) locally, in the individual halos, rather than in the cluster halo.

\item  LERG R-AGNs belonging to a given environment type (field, cluster member, BCG) have an excess of satellites out to at least 100 kpc in projected distance compared to similar radio-quiet galaxies in respective environments. This excess is on average ~50\% for field R-AGN. For many galaxies this excess results in having a massive satellite vs. not having any. This finding is similar to the \cite{2008MNRAS.384..953K} result of an overdensity of R-AGN with respect to optical, radio-quiet AGN, with the important difference that we control for the general type of environment. Thus, we find that even R-AGN BCGs have more galaxies in their vicinity than radio-quiet BCGs.
The excess is not the result of a mismatch in halo masses, but suggests a genuine relation between satellites and R-AGN triggering. 

\item Because the excess of red satellites is not accompanied by a deficit of blue satellites, we conclude that neither HERGs nor LERGs are instrumental in quenching star formation in their satellites. 

\item Similarly, since no excess of blue R-AGN satellites is observed we conclude that LERGs and HERGs rarely ($<1\%$ of cases) induce star formation in satellites.

\end{enumerate}

Our results shine new light on the scenario in which HERGs are found in sparse environments and are fueled by the accretion of cold gas, while LERGs prefer dense environments and accrete small quantities of hot halo gas. We find that the relevant environment is on small scales ($\sim100$ kpc). Thus, the fueling of LERGs either becomes more probable, or made more extensive and/or stronger by the excess of nearby satellites, which explains for example why not every BCG hosts an R-AGN, despite the dense cluster environment. Alternatively, excess satellites may trace an increased infall rate onto hosts that also feeds the gas to the black hole. We will address these possibilities in future work. Finally, our results regarding the feedback in satellite population have implications on the role of R-AGN feedback in general.

\newpage

\section{Acknowledgements}
We thank the referee for constructive comments and suggestions. This study uses data from the SDSS Archive. Funding for SDSS-III has been provided by the Alfred P. Sloan Foundation, the Participating Institutions, the National Science Foundation, and the U.S. Department of Energy Office of Science. The SDSS-III web site is http://www.sdss3.org/.

SDSS-III is managed by the Astrophysical Research Consortium for the Participating Institutions of the SDSS-III Collaboration including the University of Arizona, the Brazilian Participation Group, Brookhaven National Laboratory, University of Cambridge, Carnegie Mellon University, University of Florida, the French Participation Group, the German Participation Group, Harvard University, the Instituto de Astrofisica de Canarias, the Michigan State/Notre Dame/JINA Participation Group, Johns Hopkins University, Lawrence Berkeley National Laboratory, Max Planck Institute for Astrophysics, Max Planck Institute for Extraterrestrial Physics, New Mexico State University, New York University, Ohio State University, Pennsylvania State University, University of Portsmouth, Princeton University, the Spanish Participation Group, University of Tokyo, University of Utah, Vanderbilt University, University of Virginia, University of Washington, and Yale University.

This research made use of the ``K-corrections calculator'' service available at http://kcor.sai.msu.ru/

\clearpage

\appendix
\section{Photometric Redshifts}
\label{appa}

In our study, photometric redshifts (photo-z's) are used to reduce contamination from faint background galaxies.  A number of photometric redshifts are available in the SDSS database. The primary photometric redshifts from DR7, which are stored as \texttt{Photoz} in the DR7 Catalog Archive Server (CAS), are calculated using a technique that combines template fitting with an empirical fitting procedure to estimate the redshift \citep{2009ApJS..182..543A}. For the eighth data release, this method was also used with the inclusion of the  galaxy's inclination angle to reduce systematic bias \citep{2011ApJ...730...54Y}. These redshifts are stored in the DR8 CAS as \texttt{Photoz}. An alternative approach using the random forest technique described in \cite{2010ApJ...712..511C} was used to estimate the photometric redshifts which are stored as \texttt{PhotozRF}.

\begin{figure}[h!]
\includegraphics[width=6.1in]{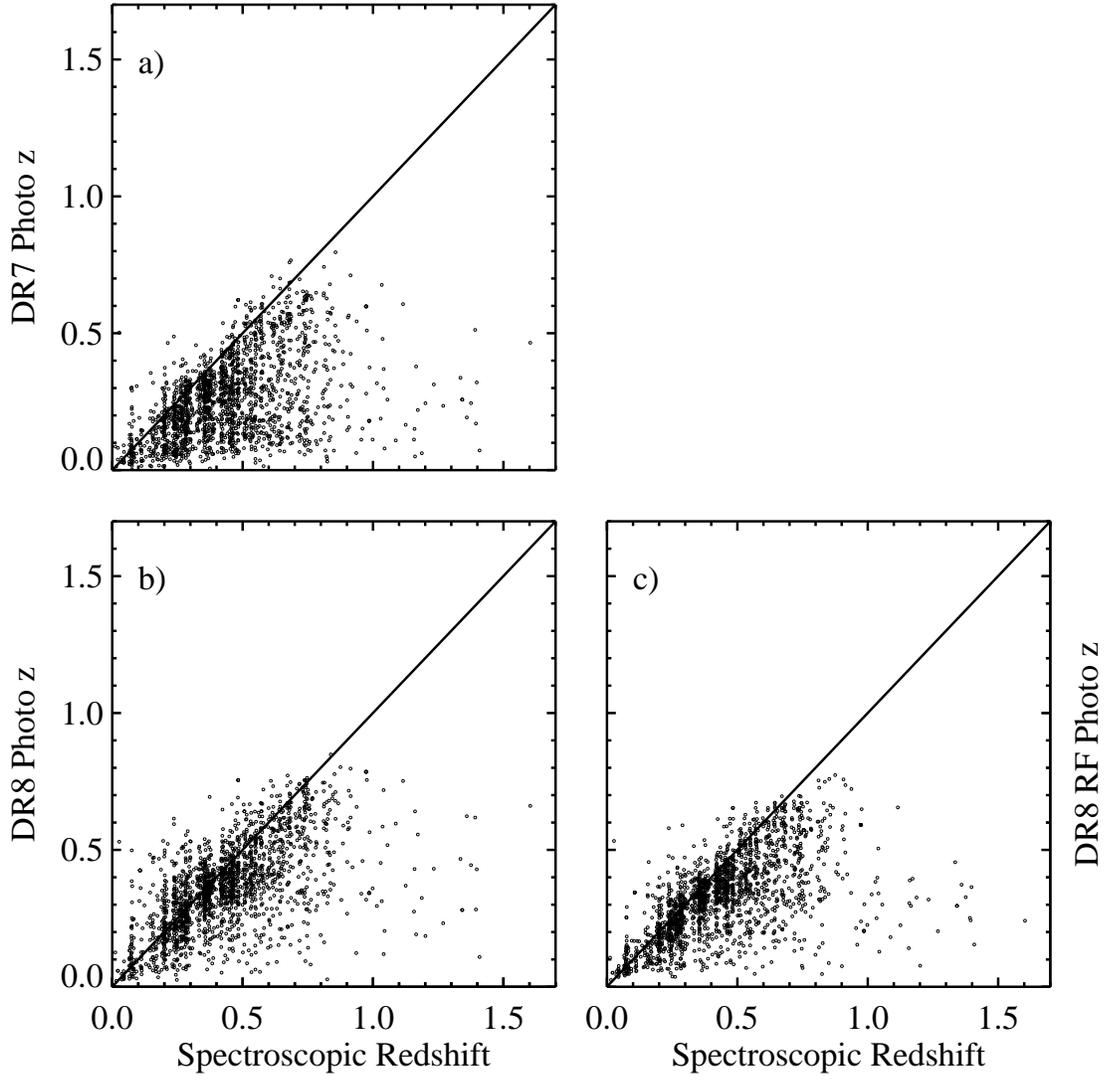}
\caption{Photometric redshifts from SDSS versus the spectroscopically determined redshift from DEEP2 (EGS field), plotted with a one-to-one line. ($a$) \texttt{Photoz} from DR7, ($b$) \texttt{Photoz} from DR8, ($c$) \texttt{PhotozRF} from DR8. Photometric redshifts from DR7 tend to be more overestimated than those from DR8.}
\label{photoz}
\end{figure}

 In order to determine which photometric redshift is the most consistently accurate, we compared the three photometric redshifts from DR7 and DR8 to the spectroscopic redshifts (spec-z's) of a sample of galaxies from the Deep Extragalactic Evolutionary Probe 2 (DEEP2) survey \citep{2007ApJ...660L...1D}. We began with the unique redshift catalog from the DEEP2 Data Release 3 (DR3), of which only galaxies with a reliable redshift and within the Extended Groth Strip (EGS) were used. These galaxies were then searched for in both DR7 and DR8, and 3,600 were found to have photometric redshifts in both data releases. We used the same magnitude cut of $r$ = 22 as was used for the galaxies from SDSS, which left 1,897 galaxies. The spectroscopic redshifts of these galaxies from DEEP2 together with the three photometric redshifts are plotted in Fig.~\ref{photoz}. Panel $a$ shows the photometric redshifts reported in DR7, panel $b$ shows the primary photometric redshifts from DR8, while panel $c$ plots the DR8 photometric redshifts derived using the random forest technique. In all panels, the spectroscopic redshift is plotted on the abscissa.

Figure~\ref{photoz} shows that for the sample as a whole, all three methods for determining photometric redshifts tend to underestimate redshifts, and this is more pronounced for DR7. The scatter between photometric and spectroscopic redshifts is large at about 0.15 dex and is not symmetric as objects at redshifts greater than $\sim$1 have systematically underestimated redshifts. Because of this large asymmetric scatter, we decided to use photometric redshifts to only remove objects with the most discrepant (large) redshifts. We used the following process to select which photometric redshift to use. To simplify the analysis we treat our target galaxies as if they were at a redshift of $z=0.15$, the mean redshift of the R-AGN sample. We then compute the fraction of galaxies in the background (spec-z $>0.15$) of this $z=0.15$ ``galaxy" that are identified as background galaxies by their photometric redshifts. The large scatter in photometric redshifts means that many galaxies with a photo-z $>0.15$ are in fact at a spec-z $<0.15$, so we add a 0.15 dex margin and consider as background galaxies those for which photo-z $>0.3$. 

Of the 1,793 galaxies with spec-z $>$ 0.15, the DR7 \texttt{Photoz} was greater than 0.3 for 663 galaxies, or 37\% of the total. For the DR8 \texttt{Photoz}, 1,241 galaxies had an estimated redshift greater than 0.3 or 69\% of the total, while the DR8 \texttt{PhotozRF} estimated a redshift greater than 0.3 for 1,106 of the galaxies, or 62\% of the total. We therefore conclude that \texttt{Photoz} from DR8 is the most useful for identifying background galaxies whose redshift is greater than the $z$ = 0.3 redshift limit of our sample of radio galaxies.

%Our goal was to determine which photometric redshift method could best be used to eliminate such background galaxies. Since the R-AGN in the sample lie at a median redshift of $z\sim0.15$,  To quantitatively compare the three methods, we computed the fraction of galaxies whose spectroscopic redshift is greater than $z$ = 0.15 that had an estimated  photometric redshift greater than $z$ = 0.3. 

%\acknowledgments

\end{document}